\documentclass{appolb}
\usepackage{graphicx}
\usepackage{xcolor}
\usepackage{authblk}
\usepackage{amssymb}
\usepackage{hyperref}
\usepackage{amsmath}
\usepackage{mciteplus}
\usepackage{makecell}
\usepackage{cellspace} 
\usepackage{soul}
\usepackage[numbers,sectionbib,sort&compress]{natbib}

\begin{document}
\title{Analytical dispersive parameterization for S-wave $\pi\pi$ and $\pi K$ scattering%
\thanks{Talk by V. Biloshytksyi, presented at the Excited QCD 2022, Giardini Naxos, Sicily, Italy, October 24-28 2022.}%
}

\author{
Volodymyr Biloshytskyi$^1$, Igor Danilkin$^1$,
Xiu-Lei Ren$^{1,2}$,

Marc Vanderhaeghen$^1$
\address{$^1$ Institut f\"ur Kernphysik \& PRISMA$^+$  Cluster of Excellence, Johannes Gutenberg Universit\"at,  D-55099 Mainz, Germany}
\address{$^2$ Helmholtz Institut Mainz, D-55099 Mainz, Germany}
}
\maketitle
\begin{abstract}
In this proceeding, we illustrate the applicability of a new parameterization of the S-wave amplitude on the example of the $\pi\pi \to \pi\pi$ and $\pi K \to \pi K $ lattice data ($m_\pi \sim 240$ MeV) from the HadSpec collaboration. The applied parameterization follows from the dispersive representation for the inverse scattering amplitude. The left-hand cut contribution is parametrized by the series in a suitably constructed conformal variable. The crucial input in the analysis is the Adler zero, whose position we extracted from the chiral perturbation theory at next-to-leading order with the uncertainties propagated from the low-energy constants.
\end{abstract}

In recent years there is significant progress in lattice QCD studies of excited hadrons
\cite{Briceno:2017max,*Shepherd:2016dni}. These studies have great potential for determining the properties of poorly known hadronic states. Proper identification of resonance parameters, however, requires the search of the S-matrix poles in the complex plane. 
Currently, very simplistic parameterizations are used in the fits, which in some cases lead to the very wide spread of the determined resonance parameters (see e.g. \cite{Briceno:2016mjc}) or even fail to find a stable solution \cite{Wilson:2019wfr}. The application of Roy-like analyses to lattice data is impossible. The approach which is based on the partial-wave dispersion relation for the direct amplitude \cite{Danilkin:2020pak,*Deineka:2022izb}
requires an extra computational cost due to numerical matrix inversion, thus complicating its implementation in fits to lattice data. The parameterization, which was proposed in \cite{Danilkin:2022cnj} is based on writing the dispersion relation for the inverse amplitude. In general, it does not give a superior representation compared to the direct representation suggested in \cite{Danilkin:2020pak,*Deineka:2022izb}. However, the advantage is its compact analytical form which is more suitable for direct numerical implementations and superior to the simple K-matrix forms, which are currently used in lattice analyses. 
In the case of zero-angular momentum scattering, it has the following form \cite{Danilkin:2022cnj} 
\begin{align}\label{DRfor1/T_finalJ0}
\left[t_0(s)\right]^{-1} =&\left[t_0(\tilde{s}_M)\right]^{-1}+\frac{s-\tilde{s}_M}{\pi} \int\limits_{L,R}\frac{d s'}{s'-\tilde{s}_M}\frac{\text{Im} \left[t_0(s')\right]^{-1}}{s'-s}+\frac{s-\tilde{s}_M}{s_A-\tilde{s}_M}\frac{g_A}{s-s_A}\nonumber \\
\simeq &  \sum_{n=0}^{\infty} C_{n}\,\omega^n(s)+R(s,\tilde{s}_M)+\frac{s-\tilde{s}_M}{s_A-\tilde{s}_M}\frac{g_A}{s-s_A}\,
\end{align}
with an arbitrary choice of the subtraction point $s_L<\tilde{s}_M<s_{th}$. 
The integral over the right-hand cut is fixed from the unitarity condition and denoted by $R(s,\tilde{s}_M)$,
\begin{eqnarray}\label{R(s,s_M)}
R(s,\tilde{s}_M)&\equiv&  \frac{s-\tilde{s}_{M}}{\pi} \int_{s_{th}}^{\infty}\frac{d s'}{s'-\tilde{s}_{M}}\frac{-\rho(s')}{s'-s}\,,\\
\rho(s)&=&\begin{cases}
\sqrt{1-4m_\pi^2/s}\,&\text{for $\pi\pi$,}\\
\sqrt{(s-(m_K+m_\pi)^2)(s-(m_K-m_\pi)^2)}/s\, &\text{for $\pi K$.}
\end{cases}
\end{eqnarray}
The Adler zero of the scattering amplitude is accounted for by the corresponding pole of the inverse scattering amplitude with the residue $g_A$.
The contribution from the left-hand cut combined with the subtraction constant is expanded in the conformal mapping
series. The variable $\omega(s)$,
\begin{eqnarray}\label{xi-1-2}
\omega(s)&=&\begin{cases}
\displaystyle
\frac{\sqrt{s}-\sqrt{s_E}}{\sqrt{s}+\sqrt{s_E}}\,&\text{for $\pi\pi$,}\\
\displaystyle
-\frac{\left(\sqrt{s}-\sqrt{s_E}\right)\left(\sqrt{s} \sqrt{s_E}+(m_K^2-m_\pi^2)\right)}{\left(\sqrt{s}+\sqrt{s_E}\right) \left(\sqrt{s} \sqrt{s_E}-(m_K^2-m_\pi^2)\right)}\,&\text{for $\pi K$.}
\end{cases}
\end{eqnarray}
maps the entire complex plane, except for the left-hand cut, into the unit circle. The expansion point $s_E$ is in general a free parameter, but in practice, it should be adjusted to maximize the convergence of the conformal series near the pole in the complex plane.

\begin{figure*}[ht]
\centering
\includegraphics[width=0.48\textwidth]{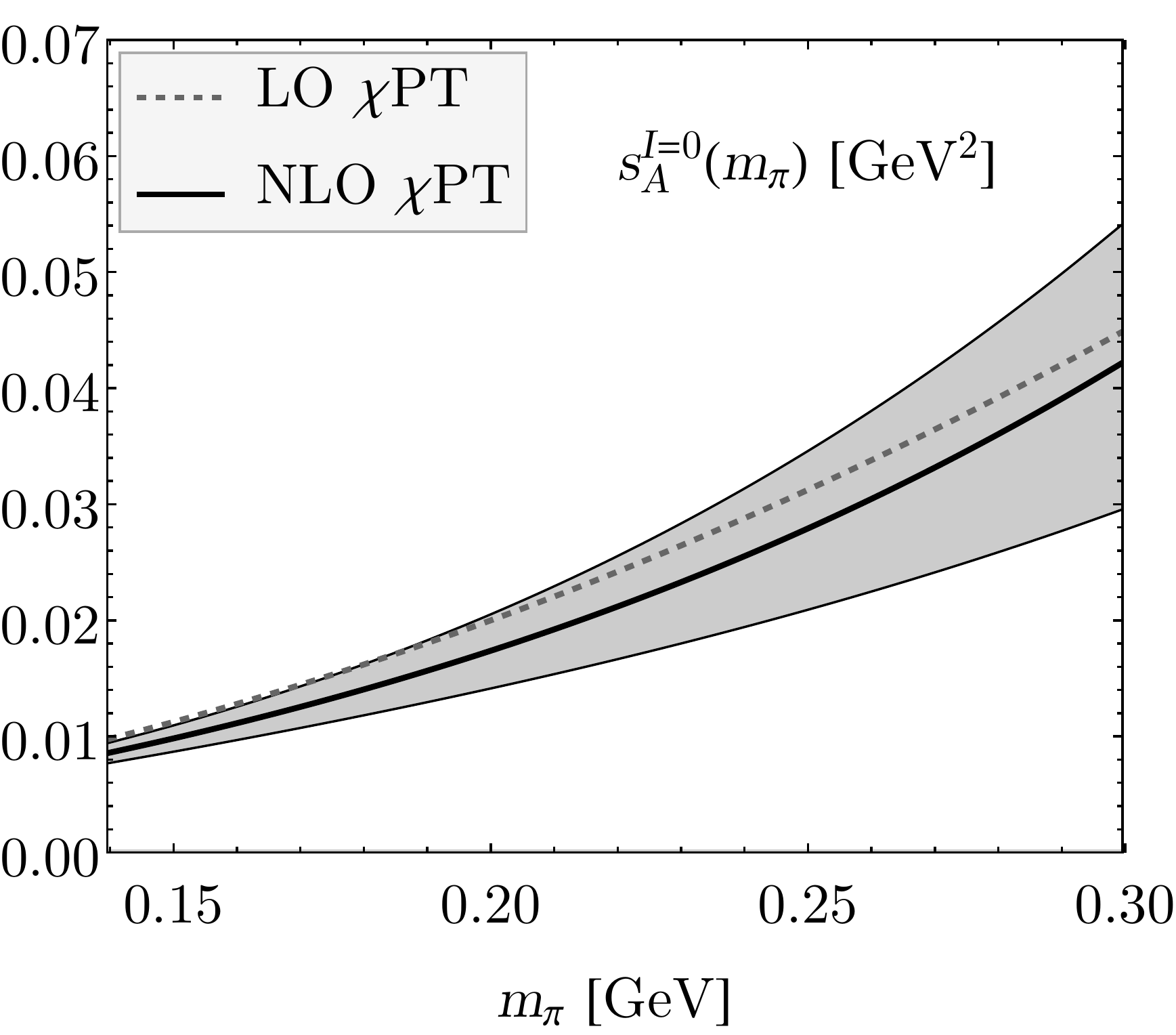}
\includegraphics[width=0.48\textwidth]{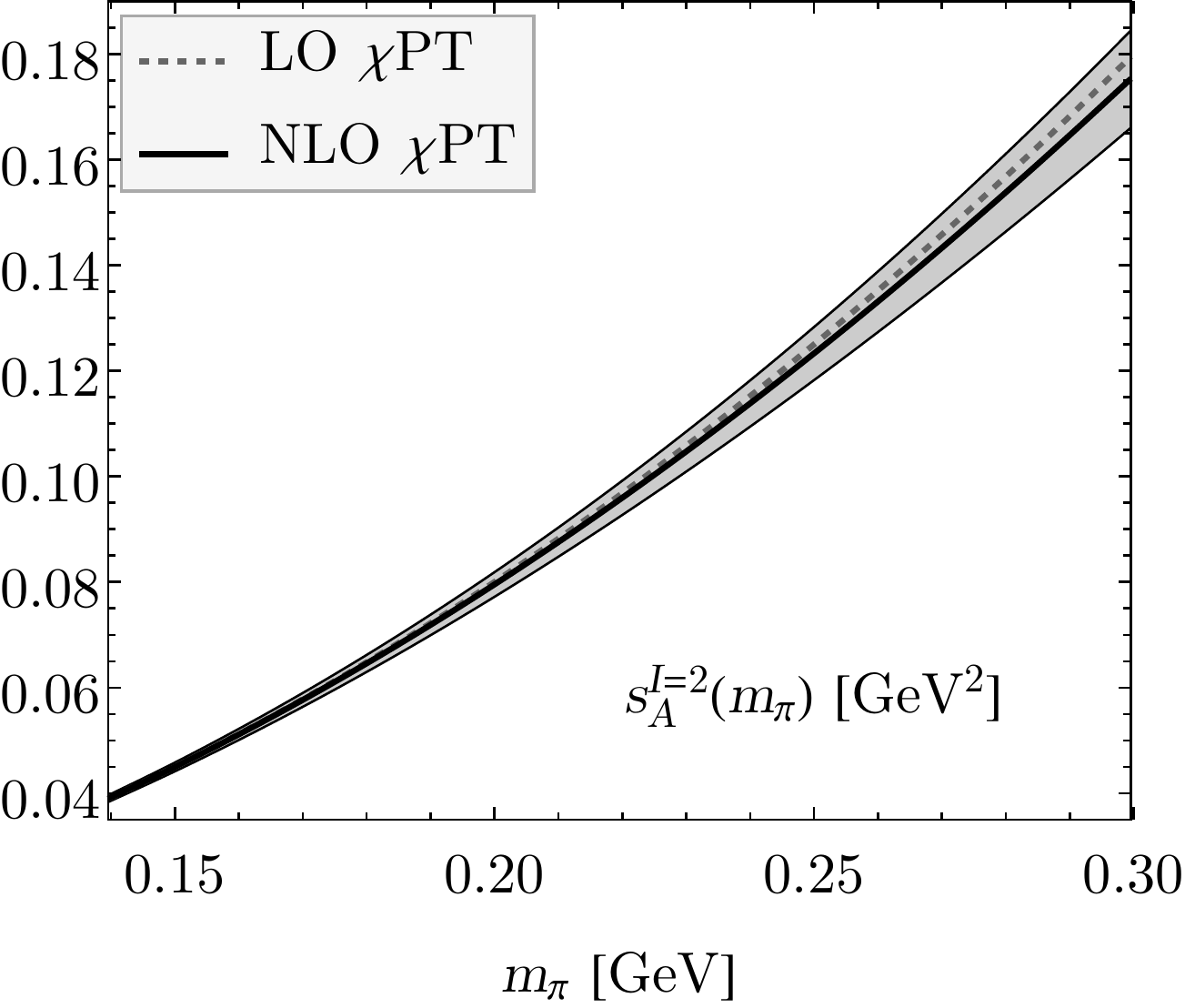}
\includegraphics[width=0.48\textwidth]{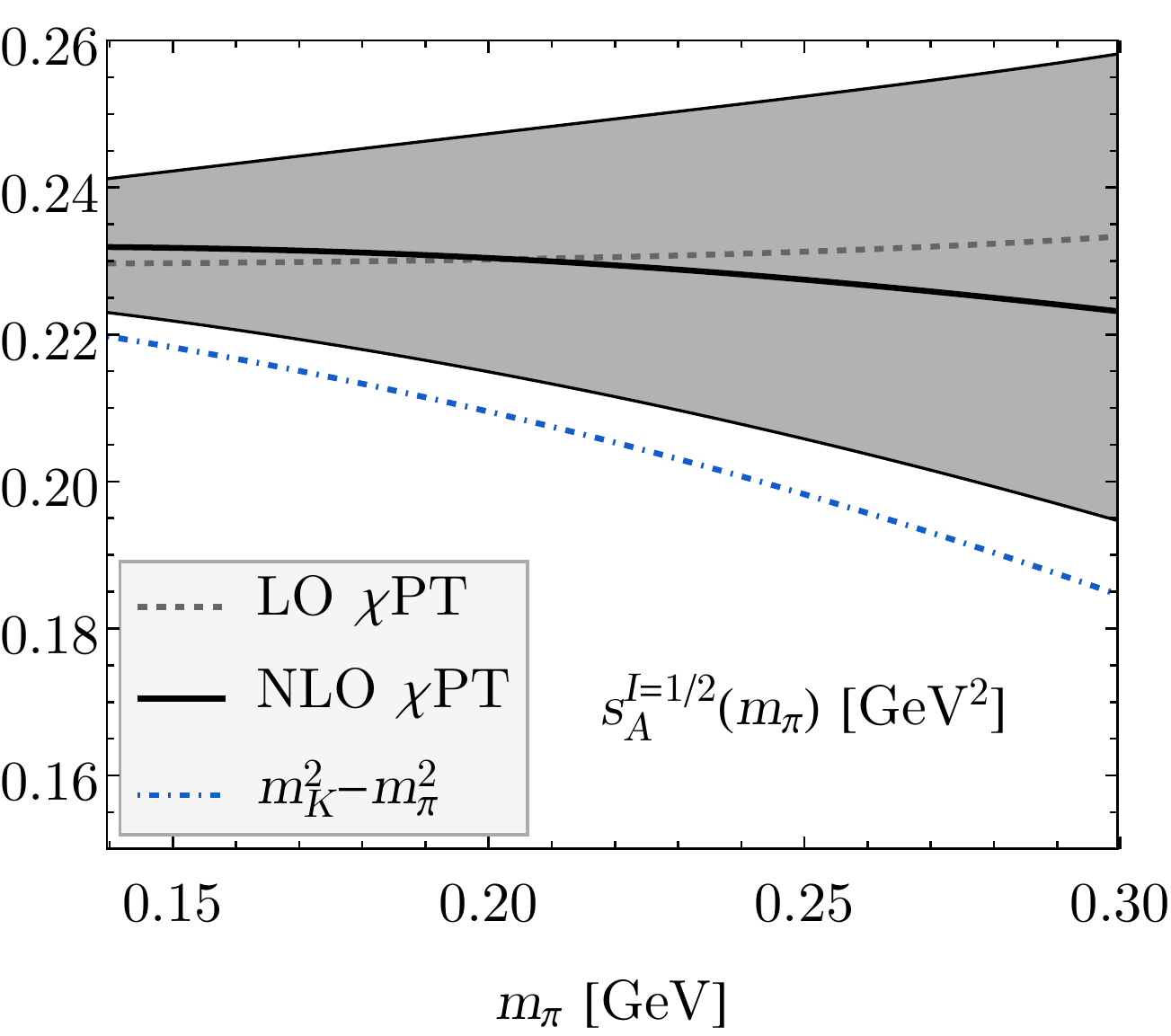}
\includegraphics[width=0.48\textwidth]{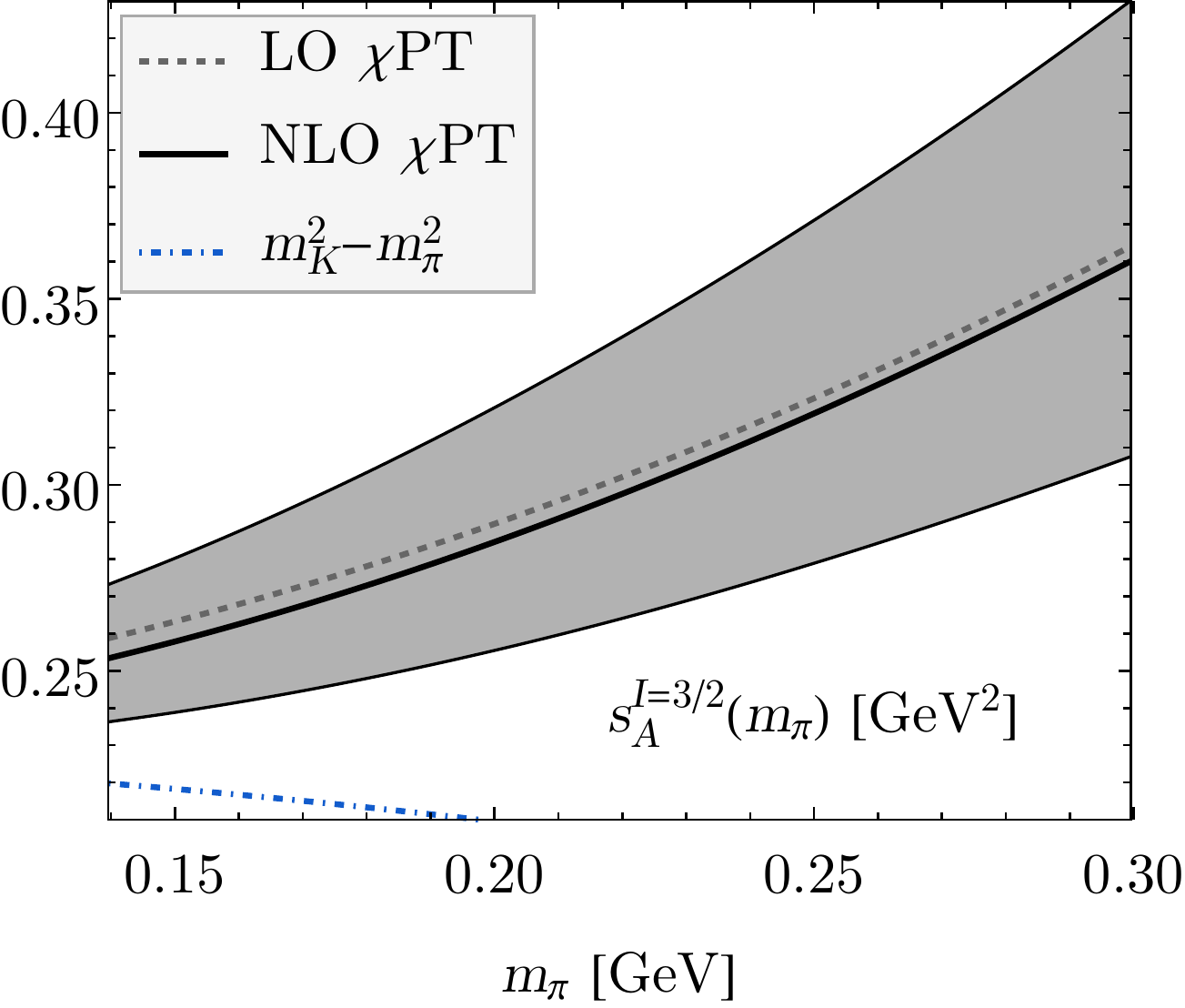}
\caption{The positions of the Adler zero, predicted by LO and NLO $\chi$PT, with respect to the pion mass. For $I=1/2,3/2$ we fixed the strange-quark mass effectively via the LO $\chi$PT meson mass relations with $m_\pi=239$ MeV and $m_K=508$ MeV taken from \cite{Wilson:2019wfr}. The gray bands illustrate the 1$\sigma$ uncertainty propagated from the LECs using the bootstrap technique. The black curves correspond to the central values of LECs.   \textit{Top panel:} $I=0,2$ from SU(2) $\chi$PT with LECs from \cite{Bijnens:2014lea}. \textit{Bottom panel:} $I=1/2,3/2$ from SU(3) $\chi$PT with LECs ("$p^4$"-fit) from \cite{Bijnens:2014lea}. The blue dot-dashed line shows the closest left-hand cut branch point.}
\label{fig:AdlerZeros}
\end{figure*}

The proposed parameterization enables the accurate description of resonances through the use of only a few terms in a conformal mapping series. 
It was explicitly demonstrated in \cite{Danilkin:2022cnj} on the well-known examples of $\pi\pi$ and $\pi K$ scattering for the physical pion mass.
However, to achieve an accurate result, the parameterization necessitates an input for the Adler zero. Given that the Adler zero resides within the domain of convergence of chiral perturbation theory ($\chi$PT) the position of the zero for the physical pion mass can be efficiently determined through the use of $\chi$PT without encountering any difficulties. The question arises for the pion mass beyond its physical value when the convergence of the $\chi$PT series becomes worse. 

\begin{figure*}[ht]
\centering
\includegraphics[width=0.5\textwidth]{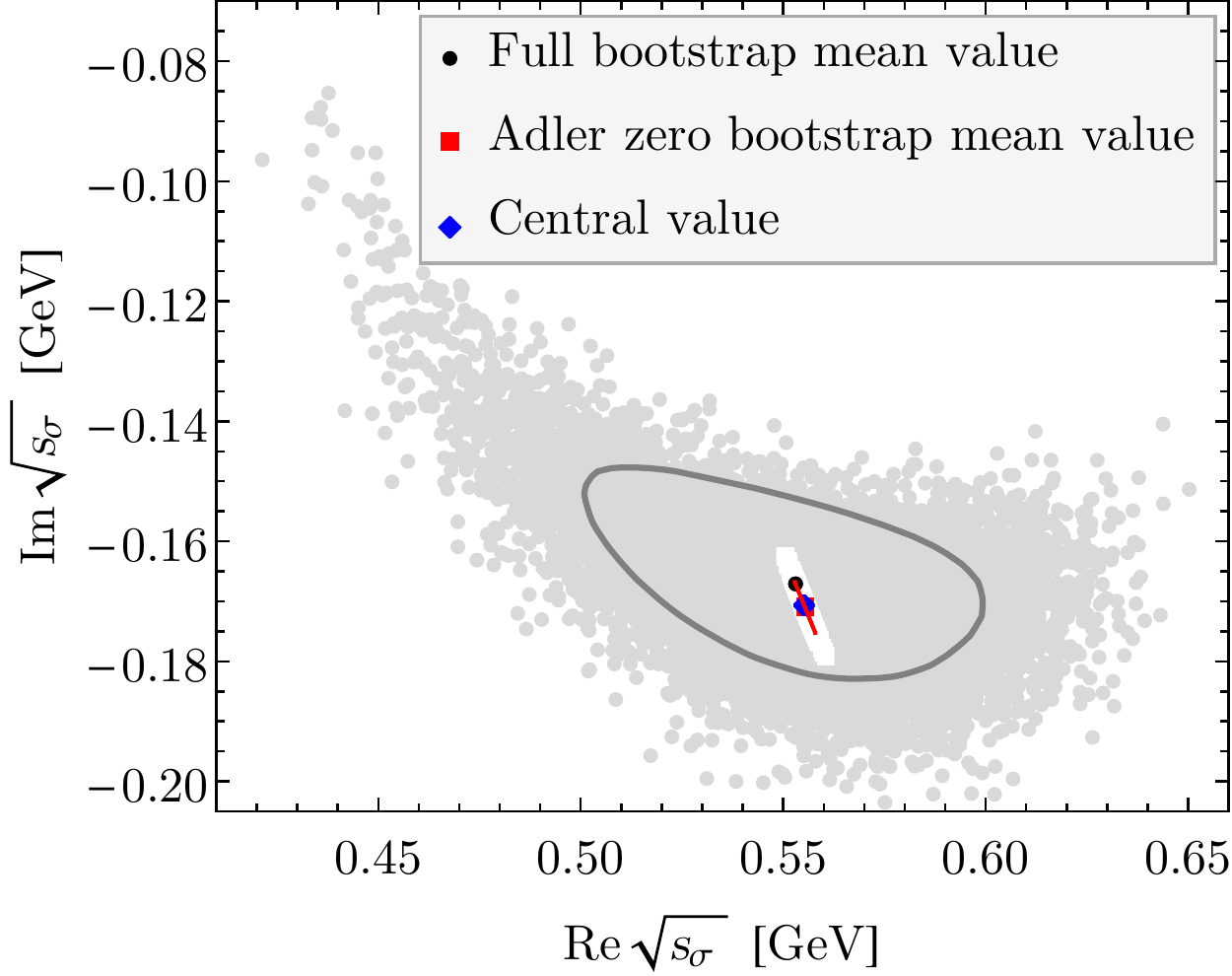}
\includegraphics[width=0.445\textwidth]{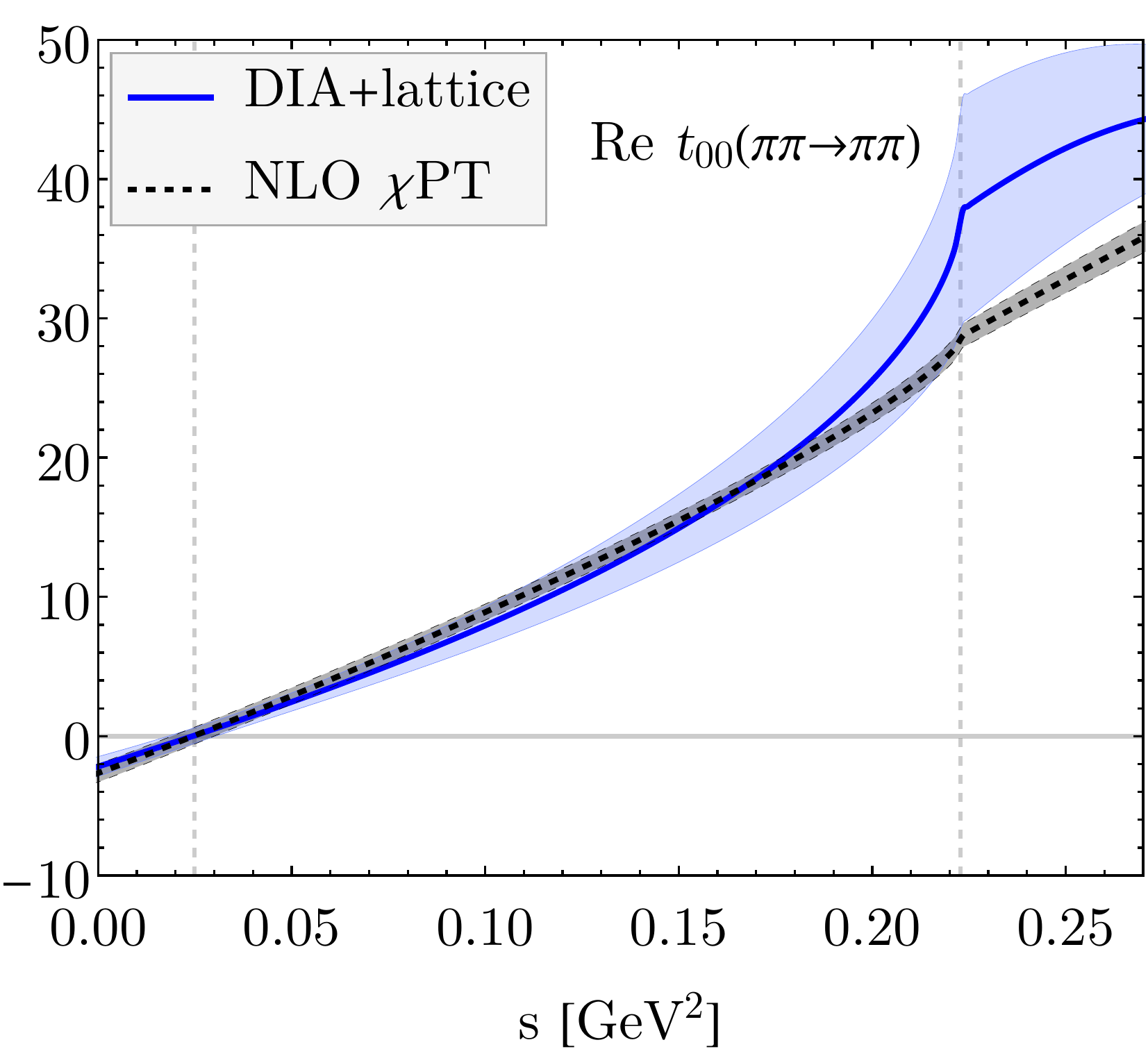}
\includegraphics[width=0.5\textwidth]{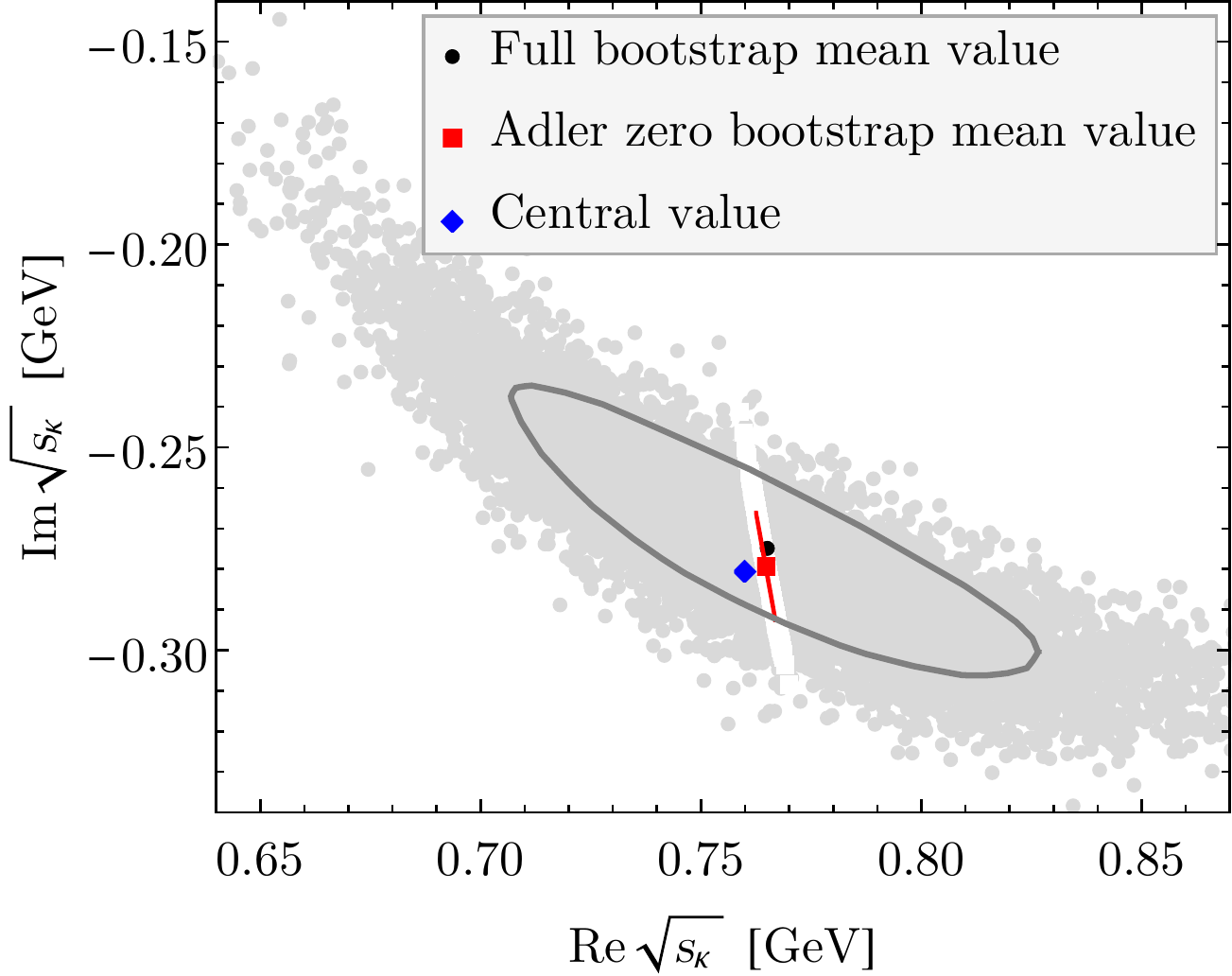}
\includegraphics[width=0.445\textwidth]{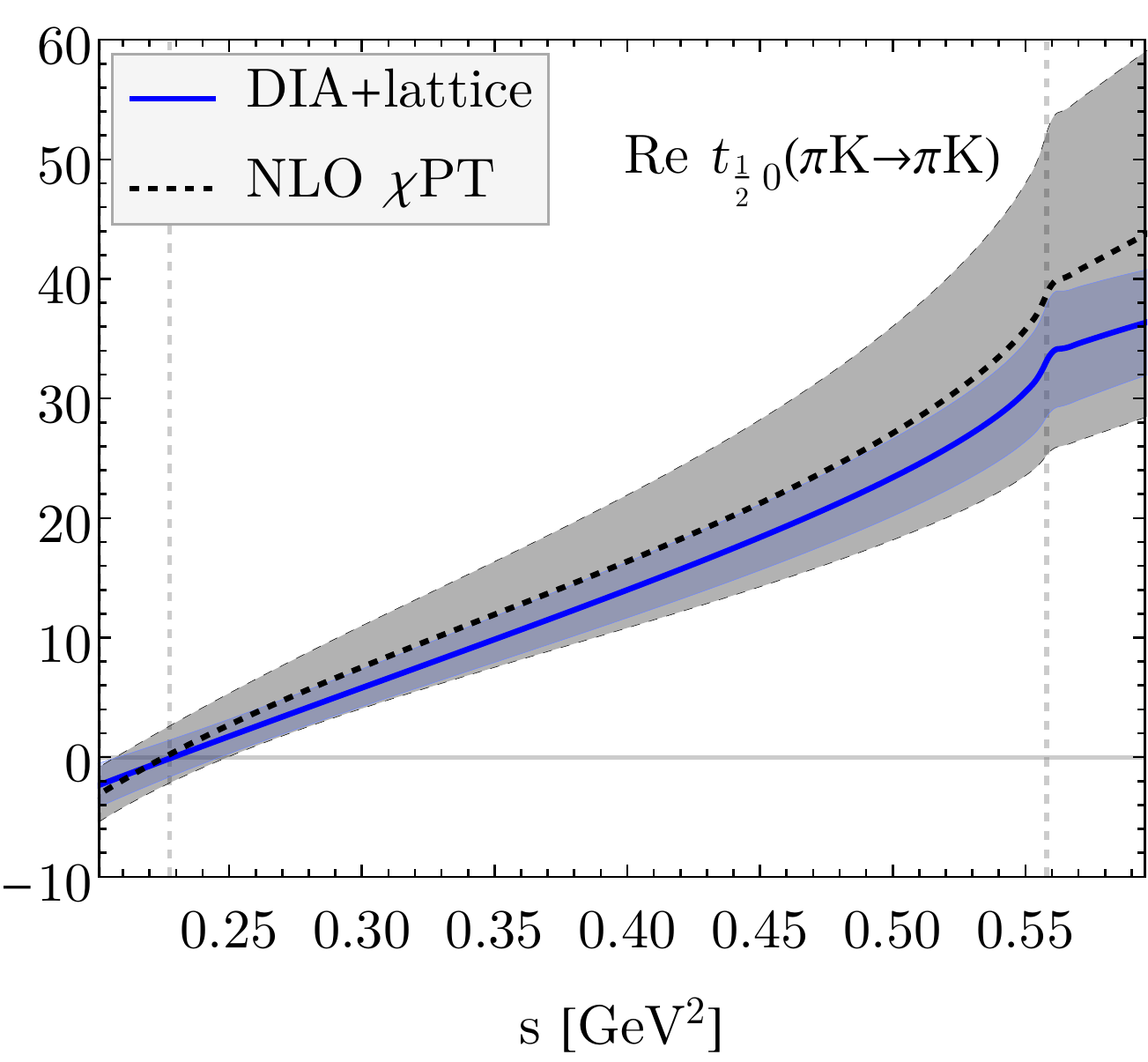}
\caption{\textit{Left panel:} the $\sigma$ and $\kappa$ poles $(m_\pi \sim 240\,\text{MeV})$ are shown with their $1\sigma$ uncertainties. The full bootstrap mean value (black point), its uncertainty (black line) and bootstrap points (gray points) are given. The mean value of the Adler zero bootstrap (red point), the corresponding uncertainties (red line) and the bootstrap points (white points) are also shown. The central value (blue point) corresponds to the pure fit without an error analysis. \textit{Right panel:} The dispersive inverse amplitudes (DIA) from the fit to the lattice data (blue bands) \cite{Briceno:2016mjc,Wilson:2019wfr} compared to the extrapolated NLO $\chi$PT results (gray bands).}
\label{fig:AmplitudesPoles}
\end{figure*}
The behavior of the Adler zero, given by $\chi$PT at LO and NLO, with the pion mass in the range from its physical value up to $300$ MeV is shown in Fig.~\ref{fig:AdlerZeros}. The error bands are estimated by propagating the uncertainties from the low-energy constants\footnote{We use those LECs that do not depend on meson masses, but on the regularization scale. Hence they are treated as fixed parameters in chiral extrapolation.}  (LECs) of the theory, assuming that they are uncorrelated. The SU(2) $\chi$PT amplitudes for $\pi\pi$ scattering are taken from \cite{Niehus:2020gmf} with the LECs from \cite{Bijnens:2014lea}. One possibility is to express the p.w. amplitude in terms of the pion decay constant in the chiral limit $f_0$, which does not depend on the pion mass. This approach has been performed in \cite{Danilkin:2022cnj}. Here we would like to follow the alternative strategy and expand the $\chi$PT amplitudes in terms of the physical pion decay constant $f_\pi$. The pion mass dependence of the latter is fixed by its NLO perturbative expansion.
The same approach we apply for the case of $\pi K$ scattering. The corresponding SU(3) $\chi$PT amplitudes are taken from \cite{GomezNicola:2001as}, with the LECs from  ``p$^4$''-fit \cite{Bijnens:2014lea}. The strategy described above effectively resumes some of the higher-order $\chi$PT corrections and enhances the $\chi$PT convergence. It results in reduced uncertainties of the Adler zero positions and improves the behaviour of the $\pi K$ partial-wave amplitudes with the isospin $I=3/2$ for the large pion masses.

\begin{table}[t]
    \centering\setcellgapes{4pt}\makegapedcells
    \begin{tabular}{c|c c}
    \hline
       Pole & Adler zero bootstrap & Full bootstrap \\
         \hline
    $\sqrt{s_\sigma}$, MeV & $555(3)-i171(4)$ &  $553^{+46}_{-52}-i167^{+19}_{-16}$\\
    $\sqrt{s_\kappa}$, MeV &  $765(2)-i279(13)$ & $765^{+61}_{-58}-i275^{+40}_{-31}$\\
    \hline
    \end{tabular}
    \caption{The parameters of $\sigma$ and $\kappa$ poles, which correspond to the left panel of Fig.~\ref{fig:AmplitudesPoles}.}
    \label{tab:PoleParameters}
\end{table}

For the description of the lattice data from \cite{Briceno:2016mjc,Wilson:2019wfr} it is enough to stay with the leading term in the conformal mapping expansion
\begin{equation}
    \left[t_0(s)\right]^{-1} \approx C_{0}+R(s,s_{\mathrm{th}})+\frac{s-s_{\mathrm{th}}}{s_A-s_{\mathrm{th}}}\frac{g_A}{s-s_A}\,,
    \label{SimpleDIA}
\end{equation}
where we have chosen $\tilde{s}_M=s_{\mathrm{th}}$. The two fit parameters are $g_A$ and $C_0$.
The pole positions deduced from the simple fits to $p\,\cot\delta$ lattice data, are shown in Fig.~\ref{fig:AmplitudesPoles} and Tab.~\ref{tab:PoleParameters}.
The mean values and their uncertainties in the full bootstrap were obtained through a combination of bootstrapping the lattice data and the Adler zero input. The mean values of the Adler zero bootstrap, along with their uncertainties, were derived by fixing the lattice data and only bootstrapping the Adler zero input. It is evident that the main source of uncertainty in the pole parameters arises from the uncertainties associated with the lattice data points, rather than from the chiral input of the Adler zero position. Comparing the fitted amplitudes with the $\chi$PT results (see right panel of Fig.~\ref{fig:AmplitudesPoles}), one can deduce that the existing lattice data on $\pi\pi$ scattering cannot provide a constraint on the LECs. However, the amplitudes fitted to the lattice data on $\pi K$ scattering lie within the $\chi$PT uncertainty bands, thus they could already constrain some of the SU(3) LECs.

The main result of this work is the stable extraction of the $\sigma$ and $\kappa$ poles from the lattice data \cite{Briceno:2016mjc,Wilson:2019wfr} through the utilization of the dispersive inverse amplitude (DIA) parameterization (Eqs.~(\ref{DRfor1/T_finalJ0}) and (\ref{SimpleDIA})). The Adler zero input with uncertainties was obtained from the extrapolated NLO $\chi$PT amplitudes and given in Fig.~\ref{fig:AdlerZeros}. 
We conclude that DIA parameterization is well-suited for the future implementations of the upcoming lattice data (see e.g. \cite{Mohler:2019talk,Rodas:2022talk}). 

\section*{Acknowledgements}
This work was supported by the Deutsche Forschungsgemeinschaft (DFG, German Research Foundation), in part through the Research Unit [Photon-photon interactions in the Standard Model and beyond, Projektnummer 458854507 - FOR 5327], and in part through the Cluster of Excellence [Precision Physics, Fundamental Interactions, and Structure of Matter] (PRISMA$^+$ EXC 2118/1) within the German Excellence Strategy (Project ID 39083149).

\bibliographystyle{apsrevM}
\bibliography{APBproc.bib}
\end{document}